\documentstyle[12pt,physrev]{article}
\renewcommand{\baselinestretch}{1.3}
\begin{document}

\begin{center}

{\Large {\bf Microscopic calculations of the spectra of light nuclei}}

\vspace{0.3in}

D. C. Zheng, B. R. Barrett, and L. Jaqua \\
\begin{small}
{\it Department of Physics, University of Arizona, Tucson, AZ 85721, USA}
\end{small}

\vspace{0.2in}

J. P. Vary \\
\begin{small}
{\it Department of Physics and Astronomy, Iowa State University, Ames,
					IA 50011, USA}

and

{\it Institute for Theoretical Physics, University of Heidelberg,
Heidelberg, Germany}

\end{small}

\vspace{0.2in}

R. J. McCarthy \\
\begin{small}
{\it Department of Physics, Kent State University, Asatabula,
Ohio 44004, USA}
\end{small}

\end{center}

\thispagestyle{empty}

\vspace{0.2in}

\begin{abstract}
We perform large-space shell-model calculations for
the low-lying energy spectra of a few light nuclei,
$^4\mbox{He}$, $^5\mbox{He}$, $^6\mbox{Li}$ and $^7\mbox{Li}$,
in a no-core model space with a realistic effective two-body
interaction (Brueckner G-matrix). Our G-matrices are calculated for the
Reid-soft-core potential in a harmonic oscillator basis.
Single-particle ``--$U$'' insertions
are replaced by two-particle ``-U'' insertions and are included
in the G matrix calculations, which sum them to all orders.
With the starting energy of the G-matrix chosen to give
approximately the experimental binding energy,
we obtain nuclear energy spectra
which are in reasonable agreement with experiment.
We also investigate the dependence of our results
on the size of the model space
and on the harmonic oscillator basis parameter ($\hbar\Omega$).

\end{abstract}

\pagebreak

\section{Introduction}
One of the major outstanding problems in nuclear physics is to calculate
nuclear properties microscopically, using a realistic nucleon-nucleon
(NN) potential \cite{bran,bk1,eo}.
Despite early successes \cite{kb},
efforts to perform such calculations in the past have
been plagued with numerous problems, the most serious being the
divergence of the perturbation-series expansion for the
shell-model effective interaction \cite{sw}. It is well-established
\cite{sw,bk2} that such an expansion is at best asymptotically convergent.
Other problems have centered around the choice of the unperturbed
basis, the treatment of spurious center-of-mass ({\sc cm}) motion and the
best NN potential to use.

In this paper we address three of these problems in an attempt to produce
meaningful, first principle results for the low-lying structure
of light nuclei. We explicitly subtract the center-of-mass ({\sc cm})
kinetic energy from the Hamiltonian,
producing a residual interaction containing the single-particle
({\sc sp}) potentials as an $A$-dependent two-body operator,
where $A$ is the atomic number of the nucleus. This
$A$-dependent two-body operator for the {\sc sp} ($U$) potential
is included in our calculation of the
Brueckner reaction matrix G \cite{gmatrix}, which sums to all orders.

Most importantly, by carrying out our calculations in large, ``no-core''
model spaces with all $A$ nucleons active, we avoid the divergence problems
for the shell-model interactions associated with core-polarization
processes \cite{nc1,nc2}.
Namely, in a no-core model space,
there are no hole states, so the terms contributing to the effective
interaction contain only particle states, such as the two-particle
ladders (or G-matrix) for two-nucleon systems (see also Ref.\cite{pb}).
For the present calculations,
we assume that three-body and higher-body effective interactions can be
neglected, so that the shell-model effective interactions can be
reasonably approximated simply by the reaction matrix G. The
price that one has to pay with a no-core model space is
that currently, due to the space and speed limitations of computers,
sensible calculations can only be done for very light nuclei.

In the present work, we consider four nuclei $^4\mbox{He}$,
$^5\mbox{He}$, $^5\mbox{Li}$ and $^7\mbox{Li}$. A shell-model study
of these and other $0p$-shell nuclei has been made
by van Hees and Glaudemans \cite{hees}, with a no-core
(0+1)$\hbar\Omega$ model space, but using an interaction determined
empirically from a fit to selected energy levels.

In section \ref{g-matrix},
we describe our G-matrix calculations which determine
our shell-model effective interactions. Our results for nuclei with
$A$=4--7 are given in section \ref{results}.
The dependence of our results on the
starting energy and the oscillator parameter is discussed in section
\ref{basis}, before our conclusions in section \ref{conc}.

\section{G-Matrix Calculation}
\label{g-matrix}
Assuming that there are only two-body interactions, we can write
the Hamiltonian for an $A$-body system as
\begin{equation}
H = \left[ \sum_{i}\frac{\mbox{\boldmath $p$}_{i}^{2}}{2m}
  - \frac{\mbox{\boldmath $P$}^2}{2mA}\right]
	+ \sum_{i<j}^{A}v_{ij} \, ,
\end{equation}
where the first term is the kinetic energy from which the contribution
of the {\sc cm} motion, $\mbox{\boldmath $P$}^2/(2mA)$, has been
subtracted. In order to use a harmonic oscillator basis,
we treat neutrons and protons as having equal mass and
we use the following equation
\begin{equation}
\sum_{i=1}^{A}\frac{1}{2}m\Omega^2 r_i^2
= \frac{2}{A} \sum_{i<j}^{A} \frac{1}{2}\frac{m}{2}\Omega^2
	(\mbox{\boldmath $r$}_{i}- \mbox{\boldmath $r$}_{j})^2
 	+\frac{1}{2}(mA)\Omega^2
		\left(\frac{\sum \mbox{\boldmath $r$}_i}{A}\right)^2
	\equiv \frac{2}{A}\sum_{i<j}^A u_{ij}
	+ \frac{1}{2}(mA)\Omega^2\mbox{\boldmath $R$}^2
\end{equation}
to rewrite the above Hamiltonian as
\begin{eqnarray}
H &=& \left[ \sum_{i=1}^A\frac{\mbox{\boldmath $p$}_{i}^{2}}{2m} +
      \sum_{i=1}^A \frac{1}{2} m \Omega^2 r_i^2 \right]
    + \sum_{i<j}^{A}v_{ij}
    - \sum_{i=1}^{A}\frac{1}{2}m\Omega^2 r_i^2
    - \frac{\mbox{\boldmath $P$}^2}{2mA} \nonumber \\
  &=& \sum_{i=1}^Ah_i + \sum_{i<j}^{A}(v_{ij} - \frac{2}{A}u_{ij}) - H_{\rm cm}
							\label{h}
\end{eqnarray}
or
\begin{equation}
\tilde{H}\equiv H + H_{\rm cm}
     =  \sum_{i}h_i + \sum_{i<j}^{A}(v_{ij} - \frac{2}{A}u_{ij})
	\equiv  H_0 + \sum_{i<j}^{A}\tilde{v}_{ij}\, ,   \label{ht}
\end{equation}
where we have defined
\begin{equation}
H_0 \equiv \sum_{i=1}^{A} h_i = \sum_{i=1}^{A}
	\left( \frac{\mbox{\boldmath $p$}_i^2}{2m} + \frac{1}{2}m\Omega^2
	\mbox{\boldmath $r$}_i^2\right) \; , \hspace{0.2in}
{\rm and} \hspace{0.2in}
\tilde{v}_{ij} \equiv v_{ij} - \frac{2}{A} u_{ij}\, .
\end{equation}
The last term in the right-hand-side of Eq.(\ref{h}) is the
the {\sc cm} harmonic oscillator Hamiltonian
\begin{equation}
H_{\rm cm} = \frac{\mbox{\boldmath $P$}^2}{2mA}
  +\frac{1}{2}(mA)\Omega^2 \mbox{\boldmath $R$}^2\, .
\end{equation}
Obviously, when the system is in the lowest {\sc cm} state,
this term contributes an overall constant, $1.5\hbar\Omega$, to
eigenenergies of all states and, therefore,
does not alter the energy spectrum. In other
words, the Hamiltonians $H$ in Eq.(\ref{h}) and $\tilde{H}$ in Eq.(\ref{ht}),
which differ by $H_{\rm cm}$, will give the same energy spectrum.

To ensure that the system will indeed be in the lowest {\sc cm} state,
the Hamiltonian that we will actually use in the shell-model calculation is
\begin{equation}
H_{\rm sm}=\tilde{H}+\lambda \left( H_{\rm cm}-\frac{3}{2}\hbar\Omega\right).
					\label{cmm}
\end{equation}
With a choice of $\lambda \gg 1$, any excited state of the {\sc cm}
will not affect the low-lying spectrum that
concerns us.

Our next step is to construct an effective
Hamiltonian for $\tilde{H}$.
The Brueckner reaction matrix \cite{gmatrix}, or G-matrix, is defined
as an infinite sum of two-particle ladder diagrams:
\begin{eqnarray}
G(\omega) &=& P \tilde{v}_{12} P + P \tilde{v}_{12} Q
	\frac{1}{\omega - Q H_0 Q} G(\omega) \nonumber \\
	&=& P \tilde{v}_{12}P + P \tilde{v}_{12} Q
	\frac{1}{\omega - Q \tilde{H} Q} Q\tilde{v}_{12} P\, ,
\end{eqnarray}
where $H_0$, $\tilde{v}_{12}$ and $\tilde{H}$ are now defined for
a two-particle system as
\begin{eqnarray}
H_0 &=& h_1 + h_2\, ,  \nonumber \\
\tilde{v}_{12} &=& v_{12} - \frac{2}{A} u_{12}\, , \nonumber \\
\tilde{H} &=& H_0 + \tilde{v}_{12}\, .
\end{eqnarray}

In this work, we calculate the G-matrix defined above,
using the exact method given by Barrett, Hewitt and McCarthy
\cite{bhm} and later modified by Vary and Yang \cite{vy}.
Note that this G-matrix is calculated for the two-body
interaction $\tilde{v}_{12}$
which includes not only $v_{12}$ but also
a two-body ``--$U$'' term, $(2/A)u_{ij}$.

The two-particle operators $P$ and $Q$, satisfying $P$+$Q$=1,
are projection operators which project into and out of the model space,
respectively. More explicitly, the operator $Q$,
also referred to as the Pauli operator, forbids the
scattering of the two particles to an intermediate state that is
inside the model space. When, for example, a no-core model space
containing 4 major shells ($0s$ through $0f$-$1p$)
is to be used, all intermediate states in the ladder diagrams
in the G-matrix calculation will have one or both particles
outside the model space, i.e., beyond the $0f$-$1p$ major shell. This
is illustrated in Fig.1.

The G-matrix depends on a parameter, the starting energy $\omega$.
In this work, we will regard the starting energy $\omega$ as an
adjustable parameter whose value will be chosen to
give a reasonable ground-state
energy of the system under consideration.

Besides the starting energy, the G-matrix also depends on the harmonic
oscillator basis, characterized by $\hbar\Omega$, the energy spacing
between the two neighboring major shells. For one nucleus ($^6\mbox{Li}$),
we will study the dependence of the spectrum on the choice of $\hbar\Omega$.

Our results will also depend on the size of the model
space. To ensure a good treatment of the {\sc cm} motion,
a rather large model space, consisting of several major shells,
needs to be used. In this work, we will investigate the
dependence of our results on the size of the model space.

It is possible, in principle, to obtain effective interactions which are
independent of $\omega$, and spectra which are independent of $H_0$.
However, the existing theories for $\omega$- and $H_0$-independent
results involve calculations of effective many-body forces
and folded diagrams summed to all orders \cite{ls,qbox}.
Recent applications to soluble models \cite{zvb} illustrate the power of such
techniques and we will expand our future efforts in this direction.

\section{Results} \label{results}
The calculations are carried out using the {\sc oxbash} shell-model
code \cite{oxbash} in a no-core model space.
The two-body matrix elements ({\sc tbme}) of the ``residual'' interaction are
simply the G-matrix elements, calculated at a value of the starting energy
$\omega$ which gives a reasonable ground-state energy.
For the free-space NN interaction $v_{12}$, we assume no Coulomb
interaction and use the Reid-soft-core potential extended to higher partial
waves \cite{rsc}.
Note that these {\sc tbme} are $A$-dependent, because
the two-body interaction that enters the G-matrix calculation is
$\tilde{v}_{12} = v_{12} - (2/A)u_{12}$.
To these {\sc tbme}, we add the {\sc tbme} of the {\sc cm}
operator $\lambda [H_{\rm cm}-(3/2)\hbar\Omega]$.
Since we have written $\tilde{H}=H_0+\sum_{i<j}^A \tilde{v}_{ij}$,
the single-particle energies are precisely
eigenenergies of the unperturbed one-body Hamiltonian $H_0$.
They are given by
\begin{equation}
\epsilon_{nlj} =\left( 2n+l+\frac{3}{2}\right) \hbar\Omega\, ,
\end{equation}
where $n$ starts from zero.

The calculations are performed in a no-core space of various sizes.
We start with the smallest possible model space ($0s$ for $^4\mbox{He}$
and $0s$+$0p$ for $^5\mbox{He}$, $^6\mbox{Li}$ and $^7\mbox{Li}$)
and increase the model space, one more major shell a time, to
a model space containing four major shells ($N$=4).
When the $N$=4 model space is used, a
non-restricted calculation is possible only for $^4\mbox{He}$.
So for the other three nuclei that we considered, a
configuration truncation is imposed, which we describe below.

In order to have a complete separation of states of one
configuration of {\sc cm} motion from states with other configurations
of {\sc cm} motion, one must include all states in the model space
which correspond to a given set of unperturbed oscillator energies. For
example, all configurations of $0\hbar\Omega$ unperturbed energy, or all
configurations of $0\hbar\Omega+1\hbar\Omega$, {\it etc.},
constitute suitable spaces.
For such model spaces and a suitably large value of $\lambda$
in Eq.(\ref{cmm}), we expect multiple copies of the intrinsic
excitation spectra whose centroids are separated by $\lambda\hbar\Omega$.
However, we find it more convenient to work with all $A$-particle
configurations constructed from a fixed number of single-particle
shells. In this case, it is straightforward to identify the subspace
through which all the unperturbed configurations
with a given maximum oscillator energy are included.
Accordingly, the number of replicas of the intrinsic spectrum with
essentially a pure {\sc cm} motion configuration is determined.
The high-lying states with mixed {\sc cm} motion configurations
are of no concern to us.

The calculated excited state energies in
$^{4}\mbox{He}$, $^{5}\mbox{He}$, $^{6}\mbox{Li}$ and $^{7}\mbox{Li}$
are given in Table I.
The results for $^{4}\mbox{He}$, $^{6}\mbox{Li}$ and $^{7}\mbox{Li}$ are
also plotted in Figs.2--4 along with the experimental data.
In the following four subsections, we discuss the results for
$^{4}\mbox{He}$, $^{5}\mbox{He}$, $^{6}\mbox{Li}$ and $^{7}\mbox{Li}$,
in this order.

\subsection{$^{4}\mbox{He}$}
This is the only nucleus for which we are able to perform
a non-restricted calculation in the $N$=4 model space.
For this nucleus, we use $\hbar\Omega$=21 MeV and
the starting energy $\omega$ is taken to be 7 MeV.
The ground-state energy in the $N$=4 calculation is -21.918 MeV,
which is close to the value given by exact calculations \cite{he4th}
with the Reid-soft-core potential \cite{rsc}
and not far from the experimental value of -28.3 MeV.
A different choice of the starting energy
will result in a different binding energy. But a change
in the starting energy that gives rise to a shift of a few MeV
in the binding energy will not change the energy spectrum in any
significant way (see discussion in section \ref{basis}).
So we have not carefully chosen the starting
energy to reproduce the result from exact calculations.

With the starting energy fixed at 7 MeV, we calculate the excitation
energies of the first excited $0^+$ state and negative-parity states.
The first excited $0^+$ state turns out to be 33.807 MeV above the ground
state, a value that is considerably larger than 20.21 MeV given by
experiment \cite{he4ex}.
This is not surprising. The second $0^+$ state in
$^4\mbox{He}$, like other low-lying excited $0^{+}$ states in doubly
or nearly doubly closed major shell nuclei, has an exotic nature.
It is an admixture
of the $1p$-$1h$ monopole, $2p$-$2h$ and perhaps $4p$-$4h$ components,
probably largely deformed \cite{zzl,toronto}.
Another possibility, due to its proximity to breakup threshold, is
that it has a much larger rms radii than the ground state.
In order to
reproduce this type of state at the right energy in a spherical basis
calculation, the model space will probably
have to be much larger than the one
that we are currently using. This is supported in Table I,
where the results for
smaller space calculations are also listed, so we can see that the
$0^+_2$ state excitation energy decreases markedly
with increasing size of model space.

The odd-parity states, on the other hand, are reproduced reasonably
well, although they are still about 1--6 MeV too high compared with
experiment \cite{he4ex}. It is remarkable that most
odd-parity states are reproduced in
a correct sequence, as they appear in the experimental spectrum.
This can better be seen in Fig.2 in which we display
the calculated and experimental spectra of $^4\mbox{He}$.

We note from Table I that although these odd-parity states are believed
to be mostly $1\hbar\Omega$ $1p$-$1h$ states and can indeed be obtained
in a small-space calculation, as had been done many years ago by
Barrett \cite{barr4}, their energies turn out to be
about 9 to 15 MeV too high in a smaller space ($0s$+$0p$) calculation.
As the size of the model space increases
from $N$=2 to $N$=4, the excitation energies
of these states decrease by 6 to 10 MeV, in much better agreement
with experiment. The agreement between our $N$=4 results and
experiment is comparable to that obtained in Ref.\cite{hees}, with
a phenomenological interaction.

We also note that the ground-state energy is a non-monotonic function
of the size of the model space. With the starting energy fixed at
one value, $\omega$=7 MeV, the
ground-state energy is --23.490 MeV for $N$=1,
--24.087 MeV for $N$=2, --21.691 MeV for $N$=3, and
--21.918 MeV for $N$=4. This may be a bit surprising, but as we
go from small spaces to larger spaces, the result are affected by
several factors. As the model space increases, the effective two-body
interaction becomes weaker; its $\omega$-dependence decreases \cite{nc1};
the intermediate-energy effects change;
the neglected effects from the effective many-body forces diminish;
and the effects from the spurious {\sc cm} motion are more accurately
treated.


\subsection{$^{5}\mbox{He}$}
For this nucleus, we will focus on the excitation energy of the
$J^{\pi}$=$1/2^-$ state. The experimental value is $4\pm 1$ MeV \cite{a5to10}.
If we regard $^{5}\mbox{He}$ as an inert
$^{4}\mbox{He}$ core plus a valence nucleon in the $0p$ shell,
the excitation energy of $J^{\pi}$=$1/2^-$ is the energy difference
between the single-particle states $0p_{1/2}$ and $0p_{3/2}$. So it is
often referred to as the single-particle splitting between the
two $0p$-shell states.

The calculation for $^5\mbox{He}$ is carried out for $\hbar\Omega$=18 MeV at
a starting energy $\omega$=18 MeV.
The smallest ($N$=2) space calculation gives a splitting
of 4.386 MeV. As we increase the model space to include the next
major shell ($1s$-$0d$), the splitting increases to 5.182 MeV.
When the $0f$-$1p$ shell is also included in the model space ($N$=4),
the splitting drops to 2.500 MeV. With $N$=4, as we mentioned before,
only a restricted calculation could be performed.
Configurations with an excitation energy
[relative to the simplest configuration $(0s)^4 (0p)^1$]
larger than $6\hbar\Omega$ are not included.
But this is not a severe truncation and should not be responsible
for the dramatic decrease in the splitting when we go from $N$=3 to
$N$=4. The similar behavior of the single-particle splitting is
also observed by Wolter {\it et al.} and is
related to the inclusion of the $1p$ orbits in the
model space \cite{millener}.

It should be mentioned that a precise measurement of
this single-particle splitting is difficult. The experimental result that
we quoted above (4 MeV) has a large error bar ($\pm 1$ MeV) \cite{a5to10}.
On the theoretical side, several successful
phenomenological effective $0p$-shell interactions
\cite{ck,wc}, obtained by fitting
the experimental data for $0p$-shell nuclei, have a very small or
even negative single-particle splitting between the $0p_{1/2}$
and $0p_{3/2}$ orbits.
However, the single-particle splitting in the
fitted interactions can change dramatically, depending on the relative
weights of the lower $0p$- and upper $0p$- shell data in the fitting
procedure. In general, a large weighting of the upper $0p$-shell data
will result in a smaller splitting \cite{millener}.

It has been predicted theoretically (see Ref.\cite{a5to10})
that a low-lying $J^{\pi}$=$1/2^+$ state
exists at an excitation energy of about 5 MeV. Positive-parity states
with $J^{\pi}$=$3/2^+$ and/or $5/2^+$ at $E_x \sim 12 MeV$ are also predicted.
The first $J^{\pi}$=$1/2^+$ state in our $N$=4 calculation turns out to be
15.891 MeV above the ground state. The
$J^{\pi}$=$3/2^+$ and $J^{\pi}$=$5/2^+$ states are 3.4 and 5.0 MeV higher.
These results do not seem to support the prediction of low-lying
positive-parity states. But these states are probably
strongly coupled to the excited $0^+$ state in $^4\mbox{He}$,
for which our calculation gives an excitation energy about 13 MeV too high.

\subsection{$^{6}\mbox{Li}$}
For this nucleus, the calculation is also carried out in three
model spaces ($N$=2, 3, 4) for $\hbar\Omega$=16 MeV, at
a starting energy $\omega$=20 MeV.
For $N$=2, the calculation is done without any restriction on
configurations. For $N$=3 and 4, however, only up to
$6\hbar\Omega$ and $4\hbar\Omega$ excitations are allowed, respectively.
The numerical results are given in Table I.
The energy spectrum obtained in the $N$=4 calculation is plotted in Fig.3.

{}From Table I, one can see that the dependence of the {\it excitation}
energies in $^6\mbox{Li}$ on the size of the model space is not so strong as
observed in $^4\mbox{He}$ and $^5\mbox{He}$, for obvious reasons.
In fact, since the low-lying excited states in $^6\mbox{Li}$ can be
loosely described as $0\hbar\Omega$ states and since $^6\mbox{Li}$
is nearly spherical, one would expect to
reproduce these states in a single-major shell calculation. Indeed,
in our $N$=2 calculation, for which only the first two major shells
are included in the model space, we already obtain
a low-lying energy spectrum that is fairly close to experiment
(Table I). This energy spectrum does not experience much change as
we increase the model space. The biggest difference in excitation energy
between $N$=2 and $N$=4 calculations is seen for the lowest $J^{\pi}$=$2^+$,
$T$=1 state (1.3 MeV).

The ground-state binding, on the other hand, decreases by about
2.5 MeV going from $N$=3 to $N$=4. As we mentioned in the discussion
for $^4\mbox{He}$, there are many factors coming into play here and
we note that the binding-energy trend in going from $N$=2 to $N$=4 is
similar in $^4\mbox{He}$ and $^6\mbox{Li}$.

\subsection{$^{7}\mbox{Li}$}
The results for $^{7}\mbox{Li}$ are also listed in Table I, for
$\hbar\Omega$=16 MeV and $\omega$=24 MeV.
Again, the $N$=2 space calculation is not restricted and
the $N$=3 and $N$=4 calculations have a configuration truncation,
as indicated in Table I.
The energy spectrum from the $N$=4 calculation is plotted in Fig.4.
All the states listed in Table I and plotted in Fig.4 have an
isospin $T$=1/2.
The second $J^{\pi}$=$7/2^-$ state is produced at an excitation energy
of 11.430 MeV, about 2 MeV too high compared with experiment.
Other low-lying states, however, are reproduced at a reasonable
excitation energy. In our calculation, we obtain
a $J^{\pi}$=$1/2^-$ state at an excitation energy of 10.540 MeV,
to which there is no corresponding experimental state.

The first excited state ($J^{\pi}$=$1/2^-$) comes out quite low
in the $N$=2 space calculation (0.069 MeV), it is slightly below
the ground state in the $N$=3 calculation (-0.022 MeV).
But in the $N$=4 calculation, it turns out to be 0.335 MeV above the
ground state, agreeing well with the experimental value of 0.478 MeV.
In Ref.\cite{hees}, this state is obtained at an excitation energy of
more than 2 MeV above the ground state, using phenomenological
effective interactions.

The excitation energy of the lowest $T$=3/2 state is
10.322 MeV in the $N$=2 calculation and 11.348 MeV in the $N$=3 calculation.
This also agrees well with the experimental value of 11.25 MeV.
We are not able to complete the $N$=4 calculation due to
computer limitations.

Again, since these states can be loosely described as $0\hbar\Omega$
states, the small space $N$=2 calculation is already a reasonable
approximation. For the $N$=2 calculation,
as in the large space $N$=4 calculation,
all but one state are reproduced in the experimentally
observed level sequence.

\section{Dependence on $\omega$ and $\hbar\Omega$}
\label{basis}
In this section, we investigate the dependence of the results on
the parameters in our calculation, namely, the starting energy $\omega$
and the harmonic oscillator basis parameter $\hbar\Omega$.

As we discussed above, the starting energy $\omega$
is chosen to give a reasonable binding energy
of the system under consideration.
The binding energy has a much stronger dependence on the
starting energy than the energy spectrum, which is our primary concern
here. That is, a slightly different choice of the starting energy
may result in a few MeV change in the binding energy,
but will not affect the energy spectrum in any significant way.

This is portrayed in Table II and in Fig.5, where we show the
ground-state energy and the energy spectrum of $^6\mbox{Li}$
for three different choices of the starting energy ($\omega$=20, 29, 38
MeV), all with a harmonic
oscillator parameter $\hbar\Omega$=18 MeV. The calculation is
completed in the $N$=4 model space allowing up to $4\hbar\Omega$
configurations. One can see in Table I that when we change
the starting energy $\omega$ from 20 MeV to 29 MeV and then to 38 MeV,
the ground-state energy changes, by about 3 MeV each time,
from -23.044 MeV to -26.044 MeV and then to -29.366 MeV.
Meanwhile, the excitation energy
for any low-lying state that we have calculated does not change by more
than 0.4 MeV in each step. So the excitation
energy spectrum is quite stable against
the change in the starting energy.

The dependence of the results on the parameter $\hbar\Omega$ of the
harmonic oscillator basis is also shown in Table II.
We give the ground-state energy and the spectrum of $^6\mbox{Li}$
for three different $\hbar\Omega$ values: 14 MeV, 16 MeV and
18 MeV. For each of these $\hbar\Omega$ values,
the starting energy is chosen to
give a comparable binding energy. From Table II and Fig.6, we see that the
energy spectra for different choices of $\hbar\Omega$ are
similar. When we go from $\hbar\Omega$=14 to $\hbar\Omega$=18,
the biggest change in the excitation energy relative to the ground state
occurs for the second $J^{\pi}$=$1^+$, $T$=0
state (about 1.2 MeV). However, the qualitative features of the spectrum
(e.g. level sequence, level spacing, {\it etc.}) are not altered by changing
$\hbar\Omega$ by about 30\%.
This change in $\hbar\Omega$ represents a 13\% change in the rms radius
of the unperturbed ground state. Thus, in a region of a physically
sensible choice of $\hbar\Omega$, we have obtained a significant account
of $\hbar\Omega$-independence with our results.

\section{Conclusions}
\label{conc}
Starting with the realistic NN Reid-soft-core potential, we have calculated
the energy spectra of light nuclei $^4\mbox{He}$, $^5\mbox{He}$,
$^6\mbox{Li}$ and $^7\mbox{Li}$. Our calculations are performed in a
no-core model space containing up to four major shells. The use of a
no-core model space not only greatly simplifies the calculation of the
shell-model effective interactions, because no diagrams with hole lines
exist; it also avoids the convergence problem
associated with summation over the perturbation expansion series.

The single-particle ``--$U$''
insertions, present when a harmonic oscillator basis is used, but often
neglected before, are included in our G-matrix which, therefore, corresponds
to a two-body residual interaction $[v_{12}-(2/A)u_{12}]$, rather than
$v_{12}$ alone. Our effective shell-model interaction, which for a no-core
space is taken to be the G-matrix,
excludes the {\sc cm} kinetic energy by explicitly
subtracting $\mbox{\boldmath $P$}^2/(2Am)$ from the Hamiltonian.
The center-of-mass
motion is forced to be in its lowest state through the addition of
$\lambda [H_{\rm cm}-(3/2)\hbar\Omega]$
to the effective interactions. This is a
feature available with the {\sc oxbash} shell-model code \cite{oxbash}.

Besides the parameter of the harmonic oscillator basis, $\hbar\Omega$,
(a small variation of which does not alter the energy spectrum obtained in
any significant way,)
our calculation has only one adjustable parameter, the starting energy
$\omega$, which is chosen to yield a reasonable nuclear binding energy.
Our results are gratifying in the sense that good low-lying
energy spectra are obtained with the same G-matrix that reasonably reproduces
the experimental binding energy.

Note that, in our no-core approach, there are no adjustable
single-particle energies to be fixed from experiment.
In effect, the single-particle properties must come from the
underlying NN interactions. In so far as the results
agree with experiment and with the phenomenologically
successful nuclear shell model \cite{talmi,bohr}, our
approach can be considered an {\it ab-initio}
method for nuclear spectroscopy.

There are some observable variations in the numerical results for both
the binding energies and energy spectra as
we go from small model spaces to large spaces, although the
qualitative feature of the results has not changed.
The need for even larger space calculations is seen as our results have
not yet stablized as the model space is increased to the limit
of our current computing capability.

Another subject of interest would be to remove the starting-energy
dependence in the calculation. As mentioned above, it has been argued that
the use of the Lee-Suzuki method \cite{ls} should enable us to obtain
a starting-energy-independent effective interaction through an iterative
procedure. However, the Lee-Suzuki approach, like many others,
involves a many-body ``Q-box'' \cite{qbox}, which is
difficult to evaluate for a system containing more than two valence particles.
Nevertheless, further work in this direction appears warranted
based on our results here.
One has a reasonable hope that the effective many-body forces need
only be evaluated approximately.

In this work, we have used the Reid-soft-core potential as the
free-space NN interaction. A variety of NN interactions exist now
and a study of the dependence of the results on the choice of the NN
potential will be of importance to the understanding of the NN potential
from the nuclear structure point of view.

\section*{Acknowledgment}
We thank D. J. Millener for helpful discussions and valuable comments.
One of us (D.C.Z.) thanks L. Zamick for useful communications.
Three of us (D.C.Z., B.R.B. and L.J.) acknowledge
partial support of this work by the National Science Foundation,
Grant No. PHY91-03011. One of us
(J.P.V.) acknowledges partial support by the U.S.
Department of Energy under Grant No. DE-FG02-87ER-40371, Division
of High Energy and Nuclear Physics and partial support from the
Alexander von Humboldt Foundation.

\vspace{0.3in}

\begin{small}

\end{small}


\pagebreak

\renewcommand{\baselinestretch}{1.0}

\begin{small}

\vspace*{-0.4in}

\noindent
{\bf Table I}. The calculated ground-state energies and excitation
energies of low-lying states, in units of MeV,
of $^4\mbox{He}$, $^5\mbox{He}$, $^6\mbox{Li}$ and $^7\mbox{Li}$.
The G-matrices are calculated for the interaction
$v_{RSC}(r_{12})-\frac{2}{A}u(r_{12})$ (see text for
more details) where $v_{RSC}(r_{12})$ is the Reid-soft-core
potential. For each nucleus, the starting energy $\omega$
is chosen to give a reasonable ground-state energy.
The calculations are performed
using the {\sc oxbash} shell-model program \cite{oxbash},
in a no-core model space of
different sizes, signified by $N$, the number of major shells contained.
Unless indicated, there is no restriction imposed on
nucleon configurations within the model space.
The experimental data are taken from Ref.\cite{he4ex} for $^4\mbox{He}$
and from \cite{a5to10} for $A>4$.
For $^4\mbox{He}$, the numbers in paratheses are taken from
Ref.\cite{isotope}.
When significant (and available),
the error bars are given for the experimental results.

\begin{center}
\begin{tabular}{cr|cc|crrr|c}\hline\hline
Nucleus & $\omega\;$ & $J^{\pi}$ & $T$ & $N$=1 & $N$=2 & $N$=3$^{a)}$
			& $N$=4$^{b)}$	& Experiment \\ \hline
$^4\mbox{He}$ & 7.0
	& $0^+_1$ & 0 & {\it-23.490}& {\it -24.087} & {\it -21.691} &
{\it -21.918} & {\it -28.296}\\
\multicolumn{2}{c|}{$\hbar\Omega$=21}
        & $0^+_2$ & 0 &       &  57.152 &  34.478 &  33.807 &  20.21 (20.1)\\
&       & $0^-_1$ & 0 &       &  30.403 &  26.742 &  22.351 &  21.01 (21.1)\\
&       & $2^-_1$ & 0 &       &  31.406 &  28.282 &  25.467 &  21.84 (22.1)\\
&       & $2^-_1$ & 1 &       &  32.159 &  29.685 &  27.236 &  23.33 (26.4)\\
&       & $1^-_1$ & 1 &       &  35.546 &  33.084 &  27.545 &  23.64 (27.5)\\
&       & $1^-_1$ & 0 &       &  39.273 &  35.943 &  30.443 &  24.25 (31.0)\\
&       & $0^-_1$ & 1 &       &  35.278 &  32.276 &  29.137 &  25.28 (29.5)\\
&       & $1^-_2$ & 1 &       &  39.298 &  37.697 &  29.741 &  25.95 (30.5)\\
								 \hline
$^5\mbox{He}$ & 18.0 &$3/2^-$&1/2&
  & {\it -22.156} & {\it -20.697} & {\it -19.682} & {\it -27.410}\\
\multicolumn{2}{c|}{$\hbar\Omega$=18}
                   &$1/2^-$&1/2&&   4.386 &   5.182 &   2.500 &$4\pm 1$\\
&                  &$1/2^+$&1/2&&  30.874 &  16.326 &  15.891 & See $^{c)}$\\
&                  &$3/2^+$&1/2&&  25.627 &  21.832 &  19.245 & See $^{c)}$\\
&                  &$5/2^+$&1/2&&  31.216 &  21.585 &  20.847 & See $^{c)}$\\
							 \hline
$^6\mbox{Li}$ & 20.0 & $1^+_1$ & 0 &
	& {\it -28.020} & {\it -28.296} & {\it -25.477} &{\it -31.996}\\
\multicolumn{2}{c|}{$\hbar\Omega$=16}
        & $3^+_1$ & 0 &       &   2.165 &   3.307 &   3.078 &  2.186\\
&       & $0^+_1$ & 1 &       &   3.036 &   3.079 &   3.313 &  3.563\\
&       & $2^+_1$ & 0 &       &   3.810 &   5.991 &   4.880 &$4.31\pm 0.022$ \\
&       & $2^+_1$ & 1 &       &   5.080 &   6.295 &   6.210 &$5.366\pm 0.015$\\
&       & $1^+_2$ & 0 &       &   7.059 &   8.166 &   7.023 &$5.65\pm 0.05$\\
								\hline
$^7\mbox{Li}$ & 24.0  &$3/2_1^-$& 1/2  &
	& {\it -36.801} & {\it -36.088} & {\it -32.642} & {\it -39.246}\\
\multicolumn{2}{c|}{$\hbar\Omega$=16}
  &$1/2_1^-$& 1/2  &       &   0.069 &  -0.022 &   0.335 &   0.478 \\
& &$7/2_1^-$& 1/2  &       &   4.114 &   5.629 &   5.512 &   4.630  \\
& &$5/2_1^-$& 1/2  &       &   5.252 &   6.933 &   7.007 &$6.68\pm 0.05$\\
& &$5/2_2^-$& 1/2  &       &   7.133 &   8.585 &   8.786 &   7.460 \\
& &$7/2_2^-$& 1/2  &       &   9.098 &  11.560 &  11.430 &$9.6\pm 0.1$\\
& &$3/2_2^-$& 1/2  &       &   8.067 &   9.446 &   9.730 &$\sim 9.9$\\
& &$1/2_2^-$& 1/2  &       &   9.442 &  11.239 &  10.540 &  (N/A)   \\
& &$3/2_1^-$& 3/2  &       &  10.322 &  11.348 &         &11.25     \\
								\hline\hline
\end{tabular}
\end{center}

\noindent
$^{a)}$ In the $N$=3 calculation for $^6\mbox{Li}$ ($^7\mbox{Li}$)
	up to $6\hbar\Omega$ ($4\hbar\Omega$) excitations are allowed.

\noindent
$^{b)}$ In the $N$=4 calculation for $^5\mbox{He}$ ($^6\mbox{Li}$,
	$^7\mbox{Li}$), up to $6\hbar\Omega$ ($4\hbar\Omega$) excitations
	are allowed.

\noindent
$^{c)}$ For $^5\mbox{He}$, a $J^{\pi}$=$1/2^+$ state at
	$E_x \sim 5 MeV$ and $J^{\pi}$=$3/2^+$ and/or $5/2^+$ states at
	$E_x \sim 12 MeV$ are predicted theoretically to exist \cite{a5to10}.

\pagebreak

\noindent
{\bf Table II}. The calculated ground-state energy and excitation energies
of low-lying excited states (in units of MeV)
of $^6\mbox{Li}$ for different choices of the starting energy $\omega$
and the harmonic oscillator basis parameter $\hbar\Omega$.
The calculations are performed in a no-core model space
containing four major shells, i.e., $N$=4.
The $0\hbar\Omega$, $2\hbar\Omega$ and $4\hbar\Omega$ configurations
are allowed.

\begin{center}
\begin{tabular}{l|rrr|rrr}\hline\hline
   & \multicolumn{3}{|c}{Dependence on $\omega^{a)}$}
   & \multicolumn{3}{|c}{Dependence on $\hbar\Omega^{b)}$} \\
States & $\omega$=20 & $\omega$=29 & $\omega$=38
       & $\hbar\Omega$=14 & $\hbar\Omega$=16 & $\hbar\Omega$=18 \\ \hline
$1^+_1\;0$ &-23.044 &-26.044 &-29.366 &-25.335 &-25.477 &-26.044\\
$3^+_1\;0$ &  2.896 &  2.937 &  2.968 &  3.214 &  3.078 &  2.937\\
$0^+_1\;1$ &  3.014 &  3.326 &  3.671 &  3.371 &  3.313 &  3.326\\
$2^+_1\;0$ &  4.863 &  5.058 &  5.271 &  4.730 &  4.880 &  5.058\\
$2^+_1\;1$ &  5.970 &  6.305 &  6.671 &  6.160 &  6.210 &  6.305\\
$1^+_1\;0$ &  7.380 &  7.667 &  7.976 &  6.765 &  7.023 &  7.667\\ \hline\hline
\end{tabular}
\end{center}

\hspace{0.6in}
$^{a)}$ $\hbar\Omega$ is fixed at 18 MeV.

\hspace{0.6in}
$^{b)}$ For each $\hbar\Omega$,
	$\omega$ is chosen to give a reasonable binding energy.

\end{small}

\pagebreak

\renewcommand{\baselinestretch}{1.2}
\section*{Figure Captions}

\noindent
{\bf Fig.1} The two-particle Pauli-operator $Q$ for a no-core model space
containing four major shells (i.e., $N$=4):
$0s$, $0p$, $1s$-$0d$ and $0f$-$1p$. Intermediate states in the ladder
diagrams cannot have both particles in the model space.

\hspace{0.05in}

\noindent
{\bf Fig.2} The calculated and experimental energy spectra of
$^4\mbox{He}$. The calculation is performed in a no-core model
space containing the first four major shells ($N$=4) with no
truncations in configurations. A harmonic oscillator basis with
$\hbar\Omega$=21 MeV is used. The experimental data are taken from
Ref.\cite{he4ex}.

\hspace{0.05in}

\noindent
{\bf Fig.3} The calculated and experimental low-lying energy spectra of
$^6\mbox{Li}$. The calculation is performed in a no-core model
space containing the first four major shells ($N$=4) with
$0\hbar\Omega$, $2\hbar\Omega$ and $4\hbar\Omega$ configurations
allowed. A harmonic oscillator basis with
$\hbar\Omega$=16 MeV is used.
The experimental data are taken from Ref.\cite{a5to10}.

\hspace{0.05in}

\noindent
{\bf Fig.4} The calculated and experimental low-lying energy spectra of
$^7\mbox{Li}$. The calculation is performed in a no-core model
space containing the first four major shells ($N$=4) with
$0\hbar\Omega$, $2\hbar\Omega$ and $4\hbar\Omega$ configurations
allowed. A harmonic oscillator basis with
$\hbar\Omega$=16 MeV is used.
The experimental data are taken from Ref.\cite{a5to10}.

\hspace{0.05in}

\noindent
{\bf Fig.5} The dependence of the low-lying energy spectrum
of $^6\mbox{Li}$ on the starting energy $\omega$. The results are
shown for three $\omega$ values: 20 MeV, 29 MeV and 38 MeV.
The calculation is performed in the $N$=4 space
allowing $0\hbar\Omega$, $2\hbar\Omega$ and $4\hbar\Omega$ configurations.
A harmonic oscillator basis with
$\hbar\Omega$=18 MeV is used.
The experimental data are taken from Ref.\cite{a5to10}.

\hspace{0.05in}

\noindent
{\bf Fig.6} The dependence of the low-lying energy spectrum
of $^6\mbox{Li}$ on the harmonic oscillator basis, characterized
by $\hbar\Omega$. The results are
shown for three $\hbar\Omega$ values: 14 MeV, 16 MeV and 18 MeV.
The calculation is performed in the $N$=4 space
with $0\hbar\Omega$, $2\hbar\Omega$ and $4\hbar\Omega$ configurations
included. For each choice of $\hbar\Omega$,
the starting energy is chosen to give a reasonable ground-state energy
(see Table II).
The experimental data are taken from Ref.\cite{a5to10}.

\end{document}